\documentclass[proceedings]{JHEP} 

\newcommand{\beq}{\begin{equation}} 
\newcommand{\eeq}{\end{equation}} 
\newcommand{\beqa}{\begin{eqnarray}} 
\newcommand{\eeqa}{\end{eqnarray}} 
\newcommand{\mx}{\left[\begin{array}} 
\newcommand{\finmx}{\end{array}\right]} 
\newcommand{\mxp}{\left(\begin{array}} 
\newcommand{\finmxp}{\end{array}\right)} 
\newcommand{\casos}{\left\{\begin{array}} 
\newcommand{\fincasos}{\end{array}\right.} 
\newcommand{\rcasos}{\left.\begin{array}} 
\newcommand{\rfincasos}{\end{array}\right\}} 
\newcommand{\nnu}{\nonumber}

\def\lsim{\ \rlap{\raise 3pt \hbox{$<$}}{\lower 3pt \hbox{$\sim$}}\ }
\def\gsim{\ \rlap{\raise 3pt \hbox{$>$}}{\lower 3pt \hbox{$\sim$}}\ }
\def\vev#1{\langle #1 \rangle}

\def\ea{{\it et. al.}}

\conference{Third Latin American Symposium on High Energy Physics}

\title{Horizontal $U(1)_H$ symmetry: a non-anomalous model}

\author{Enrico Nardi\thanks{
Based on the article \cite{Mira:2000fx} 
written in collaboration with J. M. Mira and D. A. Restrepo.}\\
        Departamento de F\'\i sica, Universidad de Antioquia\\
        A.A. 1226, Medell\'\i n, Colombia \\
        E-mail: \email{enardi@naima.udea.edu.co}}
      
      \abstract{ Spontaneously broken Abelian gauge symmetries can
        explain the fermion mass hierarchies of the minimal
        supersymmetric standard model.  In most cases it is assumed
        that the $U(1)_H$ symmetry is anomalous.  However,
        non-anomalous models are also viable and yield an interesting
        phenomenology. Cancellation of the gauge anomalies implies the
        following results: unification of leptons and down-type quarks
        Yukawa couplings is allowed at most for two generations.  The
        $\mu$ term is necessarily somewhat below the supersymmetry
        breaking scale.  The superpotential has accidental $B$ and $L$
        symmetries, and R-parity is automatically conserved in the
        supersymmetric limit.  Anomaly canncellation also implies that
        the determinant of the quark mass matrix must vanish, wich is
        possible only if $m_{\rm up}=0$.  This solves the strong $CP$
        problem and provides an unambiguous low energy test of the
        model.}

\begin{document} 

  
  One of the most successful ideas in modern particle physics is that of local
  gauge symmetries. A huge amount of data is beautifully explained in terms of
  the standard model (SM) gauge group $G_{SM}=SU(3)_C\times SU(2)_L \times
  U(1)_Y$.  Identifying this symmetry required a lot of experimental and
  theoretical efforts, since $SU(2)_L \times U(1)_Y$ is hidden and color is
  confined.  Today we understand particle interactions but we do not have any
  deep clue in understanding other elementary particle properties, like
  fermion masses and mixing angles.  The SM can only accommodate but not
  explain these data.  Another puzzle is why $CP$ is preserved by strong
  interactions to an accuracy $<10^{-9}$.  One solution is to postulate that
  one quark is massless, but within the SM there are no good justifications
  for this.  Adding supersymmetry does not provide us with any better
  understanding of these issues. In contrast, it adds new problems.  A
  bilinear coupling for the down-type and up-type Higgs superfields
  $\mu\phi_d\phi_u$ is allowed both by supersymmetry and by the gauge
  symmetry. However, phenomenology requires that $\mu$ should be close to the
  scale where these symmetries are broken. With supersymmetry, several
  operators that violate baryon ($B$) and lepton ($L$) numbers can appear.
  However, none of the effects expected from these operators has ever been
  observed.  Since a few of them can induce fast proton decay, they must be
  very suppressed or absent.

  In \cite{Mira:2000fx} we attempted to see if by insiting on the the
  gauge principle we could gain some insight into these problems.  We
  extended minimally $G_{SM}$ with a {\it non anomalous} horizontal
  Abelian $U(1)_H$ factor, and imposed the consistency conditions for
  cancellation of the gauge anomalies.  An unambiguous prediction of
  the non anomalous $U(1)_H$ is a massless up-quark.  This represents
  a crucial low energy test for our framework.  Shall future lattice
  computations rule out $m_{\rm up}=0$ \cite{Cohen:1999kk} the model
  will be disproved.
 
  The fermion mass pattern is accounted for by means of the approach
  originally suggested by Froggatt and Nielsen (FG)
  \cite{Froggatt:1979nt}.  The $U(1)_H$ symmetry forbids most of the
  fermion Yukawa couplings. The symmetry is spontaneously broken by
  the vacuum expectation value (vev) of a SM singlet field $S$, giving
  rise to a set of effective operators that couple the SM fermions to
  the electroweak Higgs field.  The hierarchy of fermion masses
  results from the dimensional hierarchy among the various higher
  order operators. This idea was recently reconsidered by several
  groups, both in the context of supersymmetry \cite{Leurer:1993wg}
  and with an anomalous local $U(1)_H$
  \cite{Ibanez:1994ig,Binetruy:1995ru,GaugeU1,Nir:1995bu,Mira:2000gg}.
  Here we study the non-anomalous case.  Our theoretical framework is
  defined by the following assumptions: {\it 1)} Supersymmetry and the
  gauge group $G_{SM}\times U(1)_H$.  {\it 2)} $U(1)_H$ is broken only
  by the vev of a field $S$ with horizontal charge $-1$.\footnote{ We
  assume that a tree level Fayet-Iliopoulus $D$-term triggers the
  breaking of $U(1)_H$ while preseving supersymmetry.}  $S$ is a SM
  singlet and is chiral under $U(1)_H$.  {\it 3)} The ratio between
  the vev $\vev{S}$ and the mass scale $M$ of the FN fields is of the
  order of the Cabibbo angle $\lambda \simeq \vev{S}/M \sim 0.2$. {\it
  4)} The only fields chiral under $U(1)_H$ and charged under $G_{SM}$
  are the minimal supersymmetric SM supermultiplets.  {\it 5)} We also
  assume $\det M^\ell \leq \det M^d\,$, as is strongly suggested by
  the measured values of the eigenvalues of the lepton ($M^\ell$) and
  down-type quark ($M^d$) mass matrices.
  
  In the following we will use the same symbol to denote a field and its
  horizontal charge.  Upon $U(1)_H$ breaking, the Yukawa couplings $Y^u\,$,
  $Y^d$ and $Y^\ell $ of the up-type and down-type quarks and of the leptons
  are generated.  They satisfy the following relations:
\begin{eqnarray}
\nnu
Y^u_{ij}&=& 
\casos{ll}
A^u_{ij}\> \lambda^{Q_i+ u_j+\phi_u}
& 
\ {\rm if } \quad 
Q_i+u_j+\phi_u\geq0,\\
0&
\ {\rm if } \quad 
Q_i+ u_j+\phi_u<0,
\fincasos
\end{eqnarray}
and similar ones for $Y^d$ and $Y^\ell$.  The zero entries arise from
holomorphy, while $A^u_{ij}$ are numerical coefficients of order $\lambda^0$
that we will often leave understood. Let us introduce the following
combinations of charges:
%
\beqa
\label{nf}
\begin{array}{lll}
  n_u=\sum_i (Q_i+ u_i), 
&& n_d=\sum_i (Q_i+ d_i),  \\
  n_Q=\sum_i Q_i, 
&& n_L=\sum_iL_i, \phantom{\Big| }   \\
  n_\ell=\sum_i (L_i+ \ell_i), 
&& n_\phi=\phi_u+\phi_d.  
\end{array}
\eeqa
After electroweak symmetry breaking, the Yukawa couplings $Y_{ij}^{u,d,\ell}$
give rise to the fermion mass matrices $M^u$, $M^d$ and $M^\ell$.  In the
absence of vanishing eigenvalues their determinants read
\begin{eqnarray}\label{detm}
\label{detmu}
\det M^u&=&\ \langle\phi_u\rangle^3
\lambda^{n_u+
3\phi_u}\>\det A^u , \\
\label{detmd}
\det M^d&=&\ \langle\phi_d\rangle^3
\lambda^{n_d+3\phi_d}\>\det A^d ,
\\
\label{detml}
\det M^\ell&=&\ \langle\phi_d\rangle^3 
\lambda^{n_\ell+3\phi_d}\>\det A^\ell . 
\end{eqnarray}
Since all the entries in $A^{u,d,\ell}$ are of order $\lambda^0$, $\det
A^{u,d,\ell}$ is of order 1. Then the size of the determinants
(\ref{detmu})-(\ref{detml}) is fixed by the horizontal charges and by the
ratio of the Higgs doublets vevs $\tan\beta =\vev{\phi_u}/\vev{\phi_d}$.

The SM Yukawa operators are invariant under a set of global $U(1)$ symmetries:
$B$, $L$, hypercharge ($Y$) and a symmetry $X$ with charges
$X(d)=X(\ell)=-X(\phi_d)$ and $X=0$ for all the other fields.  Therefore,
shifts of the horizontal charges proportional to $L$, $B$, $Y$ and $X$ do not
affect the fermion mass matrices.  In the following, we will denote as {\it
  equivalent} two sets of charges that can be transformed one into the other
by means of shifts of this kind.  Note that the superpotential term $\mu\,
\phi_u\phi_d$ (the $\mu$-term) is not invariant under $X$, and hence it can be
different for two equivalent sets.  Experimental evidences for non-vanishing
neutrino mixings \cite{Fukuda:1998mi} imply that shifts proportional to
individual lepton flavor numbers $L_a$ ($a=e,\mu,\tau$) transform between
phenomenologically {\it non equivalent} set of charges. In fact, while these
shifts do not affect the charged lepton masses, they still produce different
patterns of neutrino mixings.  In our analysis we will work with the following
linear combinations of generators: $X$, $B$, $B$-$L$, $L_\tau$-$L_\mu$,
$L_\mu$-$L_e$, and $Y$.

Since $G_{SM}\times U(1)_H$ is a local symmetry, it is mandatory to study the
consistency conditions for cancellation of the gauge anomalies.
The mixed $SU(n)^2\times U(1)_H$ anomalies, quadratic in
$SU(n)=SU(3)_C\,,SU(2)_L\,,U(1)_Y$ and linear in the horizontal charges, can
be expressed in terms of the coefficients
\beqa
\label{coefficients}
C_3&=&n_u+n_d, \nnu \\
C_2&=&n_\phi+(3n_Q+n_L),\\ 
C_1&=&n_\phi+ \hbox{$\frac83$}n_u+
\hbox{$\frac23$}n_d+2n_\ell -(3n_Q+n_L)\,. \nnu
\eeqa
The coefficient of the mixed $U(1)_Y\times U(1)_H^2$ anomaly 
quadratic in the horizontal charges reads 
\beqa 
\nnu
C^{(2)}= \phi_u^2 - \phi_d^2 + 
\sum_i \left[Q_i^2- 2 u_i^2 + d_i^2 -
L_i^2+\ell_i^2\,\right].  
\eeqa
The pure $U(1)_H^3$  and the mixed gravitational
anomalies can always be canceled by adding 
SM singlet fields with suitable charges,
and we assume they vanish. 
If the $C_n$'s in (\ref{coefficients}) 
do not vanish, the Green-Schwarz (GS) 
mechanism \cite{Green:1984sg} 
can be invoked to remove the anomalies 
by means of a $U(1)_H$ gauge shift 
of an axion  field $\eta(x)\to \eta(x) -\xi(x)\,\delta_{GS}$.  
The consistency conditions for this
cancellation read \cite{Ibanez:1993fy}  
\beq
\label{ibanez}
{C_1}/{k_1}={C_2}={C_3}=\delta_{GS},   
\eeq
where the Kac-Moody levels of the $SU(2)_L$ and 
$SU(3)_C$ gauge groups have been assumed to be unity 
and $k_1$ is the $U(1)_Y$ (arbitrary) normalization factor.
Then the weak mixing angle 
(at some large scale $\Lambda$) is  given by 
$\tan^2\theta_W=g^{\prime 2}/g^2=1/k_1\,$. 
Using  (\ref{coefficients}),  conditions  
(\ref{ibanez}) translate into 
\beq
\label{GScondition}
2(n_\phi-n_d+n_\ell) = (k_1 -\frac{5}{3})\> \delta_{GS}\,. 
\eeq
Now, one can assume that the gauge couplings unify for the canonical
value $\tan^2\theta_W = {3}/{5}\,$ \cite{Nir:1995bu}.  Then
$n_\phi=n_d-n_\ell$ is obtained.  Alternatively, one can assume that
for some reasons the l.h.s. in (\ref{GScondition}) vanishes, and thus
predict canonical gauge couplings unification \cite{Binetruy:1995ru}.
However, in the absence of a GUT symmetry the value $k_1=5/3$ is not
compelling.  Other values of $k_1$ can be in reasonable agreement with
unification at scales $\Lambda\neq \Lambda_{GUT}$ \cite{Ibanez:1993fy}
so that $n_\phi$ and $n_d-n_\ell$ are not necessarily related in any
simple way.  If $U(1)_H$ is non-anomalous (\ref{ibanez}) and
(\ref{GScondition}) still hold with $\delta_{GS}=0$, so that the
interplay with gauge couplings unification is lost. However,
$n_\phi=n_d-n_\ell$ now follows as an unavoidable consistency
condition, giving a first constraint on the permitted horizontal
charges.

Let us now study the symmetry properties of the 
coefficients  (\ref{coefficients}).
Since for each $SU(2)_L$ multiplet  
${\rm Tr}[T_3\,Y\,H]=Y\,H\,{\rm Tr}[T_3]=0$,    
the mixed electromagnetic-$U(1)_H$ anomaly 
can be expressed in terms of $C_1$ and $C_2$ as  
$C_Q = {1\over 2}(C_1+C_2)$.
Being $SU(3)_C\times U(1)_Q$  
vectorlike, it is free of $B$ and $L$ anomalies,
and then  $C_3$ and $C_Q$ must be invariant under shifts 
of the horizontal charges proportional to $B$ and $L$.   
The same is not true for $C_1$ and $C_2$ 
separately. However, the SM is free of  $B$-$L$ anomalies 
and thus $C_1$ and $C_2$ are invariant 
under the corresponding shift.   
Also $L_\tau$-$L_\mu$ and $L_\mu$-$L_e$ have vanishing 
anomalies with $G_{SM}$, so they identify two more 
possible  shifts that leave invariant the $C_n$'s.   
In the following we state the consistency 
conditions for cancellation 
of the $G_{SM}\times U(1)_H$ gauge anomalies. 

A set of horizontal charges 
$\{H\}$ is equivalent to a second set $\{H''\}$ 
for which the coefficients $C_n''$ of   
the mixed linear anomalies vanish, if and only if 
the mixed $U(1)_Q^2$-$U(1)_H$ and $SU(3)_C^2$-$U(1)_H$ 
anomaly coefficients are equal:  
\begin{equation}
\label{condition1}
C_Q-C_3 = 0\  \Longleftrightarrow  \   
C''_1=C''_2=C''_3=0. 
\end{equation}
Moreover, if for $\{H''\}$ the charge of the $\mu$ term 
$n''_\phi$ is different from zero, the 
coefficient of the quadratic anomaly 
$\widetilde C^{(2)}$ can always be set to zero: 
\beq
\label{condition2}
        n''_\phi \neq 0 
\qquad \Longrightarrow  \qquad  \widetilde C^{(2)} = 0.
\end{equation}
As it stands, this condition is sufficient but not 
necessary. However, if all the neutrinos are mixed  
at a measurable level, condition (\ref{condition2}) 
turns out to be also necessary \cite{MNardiR}. 
In the following we take $n_\phi\neq 0$ 
in the strong sense. 

To prove (\ref{condition1}) and (\ref{condition2}), 
let us start by assuming that for the initial set
$\{H\}$ 
$C_n\neq0$. Then we can shift the charges  
proportionally  to the $X$ quantum numbers. 
$H\to H+{a\over 3}X$ yields:    
\beq
\label{shift1}
C_n \to  C'_n=C_n +\alpha_n\, a\,,
\eeq
with $\alpha_3=1$, $\alpha_2=-1/3$ and $\alpha_1=+7/3$. 
We  fix  $a=-C_3/\alpha_3$ so that $C'_3=0$.
Note that the combination 
$(C_1+C_2)/(\alpha_1+\alpha_2)-C_3/\alpha_3=C_Q-C_3$  
besides being $B$ and $L$ invariant, is  
also $X$ invariant by construction.  
Now a shift proportional to $B$ can be used to 
set $C''_2=0$. Since $C_3$ is $B$ 
invariant, $C''_3=C'_3=0$.
The sum $C'_1+C'_2$ is also  $B$ invariant and thus 
$C''_1=C'_1+C'_2=2\,C'_Q$. However,   
by assumption $C'_Q=C'_3\>(=0)$ and then the set $\{H''\}$ 
has vanishing mixed linear anomalies.  
Now, in order to cancel the quadratic anomaly
while keeping vanishing $C''_n$, 
we can use any of the  SM anomaly free 
symmetries   $B$-$L\,$,  $L_\tau$-$L_\mu\,,$  
$L_\mu$-$L_e\,$.  Since  
$L_\tau$-$L_\mu$ and $L_\mu$-$L_e$ transform between 
non equivalent set of charges, we keep this freedom 
to account for two neutrino mixings 
(the third one results as a prediction) and     
we use $B$-$L$.
Under the charge redefinition $H \to H + \beta\,$($B$-$L$) 
\beqa
{C^{(2)}}''\to  \widetilde C^{(2)} &=&
{C^{(2)}}''+ \beta\, \left[
{4\over 3} n''_u - 
{2\over 3} n''_d + 
2 n''_\ell \right] \nnu \\
&=&  
{C^{(2)}}'' - 2\,\beta\,n''_\phi,  
\eeqa
where in the last step we have used the identity 
$ {4\over 3} n_u - {2\over 3} n_d + 
2 n_\ell =  C_1+C_2 - {4\over 3}\,C_3- 2\,n_\phi$
and the vanishing of the $C''_n$. 
If,  as we have assumed, $n''_\phi\neq 0$,  
we can always set $\widetilde C^{(2)}=0$ by choosing   
$\beta=C^{(2)\prime\prime}/ (2\,n''_\phi)$.  
It is also useful to note that a set of horizontal 
charges $\{H\}$ for which $C_n=C^{(2)}=0$ identifies 
a one parameter family of anomaly free charges generated by 
shifts proportional to hypercharge: $H \to H+yY$. 
For the $C_n$'s this is trivial
due to the vanishing of the SM anomalies  
${\rm Tr}[SU(n)^2Y]=0$. For $C^{(2)}$ we have 
${\rm Tr}[Y\,H^2] \to {\rm Tr}[Y\,(H+Y)^2] = 2 C_1 =0 $. 

In summary, cancellation of the 
$G_{SM}\times U(1)_H$ gauge anomalies implies  
the following constraints on the fields charges 
%
\begin{equation}
\label{mucharge}
n_\phi \neq 0,   \qquad 
n_\phi = n_d-n_\ell 
\simeq \ln_\lambda {\det M^d \over \det 
 M^\ell}\,,   
\end{equation}
%
where the last relation follows from 
(\ref{detmd}) and (\ref{detml}).
Since $n_d\neq n_\ell$ we conclude that 
Yukawa coupling unification is permitted at most 
for two families. 
Together with assumption {\it 5)},   
we also obtain  $n_\phi<0\,$, so that 
the  superpotential $\mu$ term is forbidden 
by holomorphy and vanishes in the supersymmetric limit. 
Let us confront these results with phenomenology. 
To a good approximation the mass ratios 
$m_e/m_\mu \sim \lambda^3$, $m_\mu/m_\tau \sim \lambda^2$, 
$m_d/m_s \sim \lambda^2$ and $m_s/m_b \sim \lambda^2$
are renormalization group invariant. Then, since  
Yukawa coupling unification works remarkably well
for the third family, 
$\det M^\ell / \det M^d \sim \lambda$  and 
the preferred value of $n_\phi$  
is $-1\,$.  
Then a $\mu$ term arising from the (non-holomorphic) 
K\"ahler potential \cite{GMasiero} will have a value  
somewhat below the supersymmetry breaking scale $m_{3/2}$: 
\beq
\mu \sim \lambda^{|n_\phi|}\> m_{3/2}\qquad {\rm with} 
\qquad n_\phi=-1\,.    
\eeq
As we have explicitly shown, the  anomaly cancellation 
condition $C_Q-C_3=0$  (\ref{condition1}) is $Y$, 
$B$, $L$ and $X$  invariant and hence 
it shares the same invariance of the Yukawa couplings.   
Therefore, any product of the determinants 
(\ref{detmu})-(\ref{detml}) for which the  
overall horizontal charge  can be recasted just 
in terms of the $C_n$'s must depend 
precisely on this  combination.
Such  a relation was first found in \cite{Nir:1995bu}.  
Given that $C_Q-C_3=n_\ell-{2\over 3}n_d+{1\over 3}n_u+n_\phi$
we can write it down at once:   
\beqa 
 \label{combination}\!\!
\left({{\rm det} M^\ell\over \vev{\phi_d}^3}\right)\!\!
\left({{\rm det} M^d\over \vev{\phi_d}^3}\right)^{\!\!\!{-{2\over3}}}\!\!\!
\left({{\rm det} M^u\over \vev{\phi_u}^3}\right)^{\!\!\!{1\over3}}\!\!\!\!
\simeq\! \lambda^{C_Q-C_3}.
\eeqa
Let us confront this relation with phenomenology. 
Anomaly cancellation implies that the r.h.s. 
is unity, while the  l.h.s. is bounded by the upper 
limit 
$\left[
\left({\rm det}M^d/ \vev{\phi_d}^3\right) 
\left({\rm det}M^u/\vev{\phi_u}^3\right)
\right]^{1/3}\!\ll\! 1\,$.  
This inconsistency (or similar ones) led several authors to conclude
that $U(1)_H$ must be anomalous
\cite{Ibanez:1994ig,Binetruy:1995ru,GaugeU1,Nir:1995bu}.  However,
(\ref{combination}) is meaningful only under the assumption that none
of the determinants vanishes, and since low energy phenomenology is
still compatible with a massless up quark \cite{Cohen:1999kk,mup0old}
(see however \cite{mup0Leutwyler}) this might not be the case.  In the
following we show that insisting on the vanishing of the gauge
anomalies yields $m_{\rm up}=0$ as a prediction.

We start by noticing that if the determinant 
of the matrix $U_{ij}\!\sim\!\lambda^{Q_i+ 
u_j+\phi_u}$ has an overall negative charge 
$ \eta^U  \equiv  n_u  +  3 \phi_u $ $\sim 
\log_\lambda\det U <0 $,  
then $M^u$ has vanishing eigenvalues. In fact  
$\det U$ consists of the sum 
of six terms of the form 
$\lambda^{n_1}\cdot\lambda^{n_2}\cdot\lambda^{n_3}$  
where $n_1+n_2+n_3= \eta^U<0$. Then 
at least one of the $n_i$ must be negative,    
corresponding to a holomorphic zero in the mass
matrix. Hence each one of the six terms vanishes.  

Now, if $U(1)_H$ is anomaly free and assumption {\it 5)} holds, it is
easy to see that the determinant of the six quark mass matrix ${\cal
M}_q$ vanishes:
%
\beqa
\label{theorem2}
\rcasos{l} 
n_\ell \geq n_d  \    \\
C_n=C^{(2)}= 0 \   
\rfincasos
\  \Longrightarrow  \   \det {\cal M}_q=0. 
\eeqa
%
Adding and subtracting $3n_\phi$ 
to $C_3=0$ yields   
\beq
 \eta^U +  \eta^D = 3\, n_\phi < 0. 
\eeq
Then at least one of the two  $\eta$  
must be negative, and the corresponding 
determinant vanishes. 
Of course, on phenomenological grounds, 
a massless up quark is the only 
viable possibility \cite{Cohen:1999kk,mup0old}. 
Using the $d$-quark mass ratios
given above, and taking  
$m_b/m_t\sim \lambda^3$ 
and $n_\phi=-1$, we obtain 
\beq
\label{etau}
 \eta^U   
\simeq   -9 - 3\,\log_\lambda\left(\frac{m_b}{m_t} \tan\beta\right),  
\eeq
that ranges between $-9$ and $-18$ for $\tan\beta$ between $m_t/m_b$
and 1.  Because of the constraints from holomorphy, $\eta^U<0$
results in an accidental $U(1)_u$ symmetry acting on the $SU(2)_L$
singlet up-quark: $u_1\!\! \to e^{i\alpha} u_1$.  By means of this
chiral transformation the QCD $CP$ violating parameter
$\bar\theta\equiv \theta +{\rm arg}\det{\cal M}_q$ can be rotated
away, and is no more physical.  However, holomorphy plays a crucial
role to obtain this result, and we must check if  
it is mantained after supersymmetry breaking.
While Yukawa couplings redefinition needed
to bring the kinetic terms into canonical form cannot 
lift zero eigenvalues \cite{Mira:2000fx},   
general soft supersymmetry breaking terms do 
not respect the $U(1)_u$ symmetry, and can 
induce $m_{\rm up}\! \neq\!  0$ 
radiatively.  A conservative estimate gives
\cite{Mira:2000fx} $m_{\rm up} \lsim (\alpha_s/\pi)\,
\lambda^{|\eta^U-4|}\> \vev{\phi_u} \lsim 10^{-6}\,, (10)\,$eV [for
$\tan\beta\sim 1\,, (m_t/m_b)$]. 
The resulting contribution to
the neutron electric dipole moment is 
\cite{Pospelov:1999mv}: 
$d_n/e\lsim 10^{-28}\,\bar\theta\>,(10^{-22}\,\bar\theta)\>$cm.
Therefore, for moderate values of $\tan\beta\,$, the neutron dipole
moment remains safely below the experimental limit $d_n/e < 6.3 \times
10^{-26}\>$cm \cite{Harris:1999jx} even for $\bar \theta\sim 1$.

Gauge symmetry and supersymmetry, together with constraints from
fermion charges relations, also imply that the superpotential has
accidental $B$ and $L$ symmetries.  This result is deeply related to
the solutions of the $\mu$ and of the strong CP problems ($n_\phi<0$,
$\eta^U<0$).  The proof requires certain phenomenological inputs, like
fermion mass ratios and CKM mixings, and the assumption that neutrinos
mixings are sizeable.  Since it is somewhat lengthy, it will be
presented elsewhere \cite{MNardiR}.  An intuitive argument goes as
follows: given a set of minimal charges that fit well the fermion
masses and mixings, $\eta^U<0$ (\ref{etau}) also implies that
$C^{(2)}$ is negative.  To cancel $C^{(2)}$ the shift $H\to
H+\beta\cdot$($B$-$L$) is required, where $\beta=C^{(2)}/2n_\phi$ is
positive.  All the R-parity violating operators $\mu_L L \phi_u$,
$\lambda LL \ell\,$, $\lambda'LQ d\,$ and $\lambda'' u d d\,$ have
$B$-$L$$=-1$, so that under this shift their charges are driven to
negative values implying that they cannot appear in the
superpotential.  Of course, dimension five see-saw operators for
neutrino masses are also forbidden.  However, the same mechanism that
generates $\mu$ will generate (with larger suppressions) also $\mu_L
L\phi_u$ terms, that induce s-neutrinos vevs. Canonical
diagonalization of $L$-$\phi_d$ mixed kinetic terms will produce tiny
$\lambda$ and $\lambda'$ from the Yukawa couplings. Both these effects
can result in small neutrino masses \cite{Mira:2000gg}. However, since
none of the $\lambda''$ can be generated in this way, proton stability
is not in jeopardy.  Finally, we stress that except for $\eta^U<0$ the
condition $C_Q-C_3=0$ does not imply other serious constraints on
charge assignments.  Within our framework the mass matrices of popular
models \cite{Leurer:1993wg,GaugeU1} can be easily reproduced and,
apart from $m_{\rm up}=0$, also the same phenomenology \cite{MNardiR}.



\end{document}